\documentclass[aps,prd,showpacs,preprintnumbers]{revtex4}
\usepackage{graphics}
\def \beq{\begin{equation}}
\def \eeq{\end{equation}}
\def \beqa{\begin{eqnarray}}
\def \eeqa{\end{eqnarray}}
\def \cs{c_s}

\def \etal{{\sl et al.\/}}
\def \jhep{{\sl J.\ H.\ E.\ P.\/}}

\def \pl{{\sl Phys.\ Lett.\/}}
\def \pr{{\sl Phys.\ Rev.\/}}
\def \prl{{\sl Phys.\ Rev.\ Lett.\/}}
\def \pram{{\sl Pramana\/}}
\def \ie{{\sl i.\ e.\/}}

\begin{document}
 
\title{Dissipation, hydrodynamics and the fireball}
\author{Sourendu \surname{Gupta}}
\email{sgupta@theory.tifr.res.in}
\affiliation{Department of Theoretical Physics, Tata Institute of Fundamental
         Research,\\ Homi Bhabha Road, Mumbai 400005, India.}

\begin{abstract}
We investigate the hydrodynamics of the QCD plasma using dimensionless
numbers built from the thermodynamics and transport theory of the
plasma and characteristic dimensions of the fireball produced in
heavy-ion collisions. We find that by the usual measures, dissipation
is strong, and the fireball is on the borderline of equilibrium.
As a result, the system is richer in phenomena than ideal hydrodynamics
would predict. One general implication is that it may be possible
to get a direct view of the QCD plasma phase rather than having to
infer its existence indirectly from signals that come from the
freezeout isotherm after the fireball has cooled into the hadronic
phase.
\end{abstract}
\pacs{\hfill
TIFR/TH/05-28, hep-lat/0507210}
\maketitle

There is an emerging concordance of the first results from AdS/CFT
estimates \cite{son}, lattice gauge theory \cite{sgupta,nakamura},
and analysis of RHIC data \cite{teaney} that the QCD plasma has
small transport coefficients, \ie, small mean free times.  One of
the transport coefficients that has been measured on the lattice
is the electrical conductivity \cite{sgupta}.  This is much smaller
than weak coupling estimates would predict, implying that the
resistivity is larger. Large resistivity means that if an external
potential difference is established across the plasma, it would not
drive large currents, but instead generate large amounts of entropy.

In a similiar vein, if the viscosity of the fluid is small, then
the implication is that Reynolds numbers are large (all other
conditions being equal). Large Reynolds numbers are typically
associated with turbulence and other sources of entropy production
when the fluid is subjected to external forces. On the face of it,
this seems to be in contradiction to statements that the quark-gluon
fluid is an ``ideal liquid'', so it pays to examine this matter
more closely.  We do so, and find that the situation is complex and
interesting--- turbulence is unlikely to occur, but the fireball
is on the borderline of dis-equilibrium. This conclusion is driven
in equal measure by analysis of the data that the RHIC has produced
\cite{teaney}, the understanding of QCD gleaned through lattice
computations \cite{sgupta,nakamura}, and limits from AdS/CFT
\cite{son}. More detailed conclusions are enumerated at the end of
the article.

One distinction between a liquid and a gas that seems to be easy
to generalize is that in a gas the mean free path is much longer
than the interparticle spacing, whereas in a liquid the two are of
the same order. The number density should be replaced by the entropy
density, $s$, in a relativistic system, so one should examine the
dimensionless quantity called the liquidity in \cite{iitk}---
\beq
   \ell = \tau s^{1/3}
\label{liquid}\eeq
where $\tau$ is the mean free time. A little above the QCD cross-over
temperature, $T_c$, one may use lattice measurements of the two
quantities on the right to estimate this dimensionless number. Using
the inferred value of $\tau=0.3$ fm from lattice computations at
$2T_c$ \cite{sgupta}, and $s/T^3=5.7\pm0.2$ \cite{swagato}, one
finds that $\ell=0.8$, \cite{iitk,kol05}.  At much higher temperatures,
where weak coupling estimates \cite{amy} become reliable, one has
$s/T^3=6\pi^2+{\cal O}(g^2)$, and $\tau={\cal O}(1/g^4\log1/g)$,
so that $\ell\propto 1/g^4\log(1/g)$ when $T$ becomes sufficiently
large.  There is no phase transition between the strong coupling
regime near $T_c$ (liquid) and weak coupling (gas), so the nature
of this ``liquid'' bears further investigation.

If one takes the first order dissipative hydrodynamics of \cite{ll}
and writes it in the Eckart frame \cite{eckart}, one gets the Navier-Stokes'
equation with a relativistic correction term. Analysis of dissipation
through the Reynolds number then is a reasonable starting point.
The Reynolds number is
\beq
   R = \frac{\epsilon vr}\eta  = \frac{(\epsilon/sT)v(rT)}{\eta/s}
\label{reynolds}\eeq
where $\epsilon$ is the energy density of the QCD matter, $s$ its
entropy density, $T$ its temperature, $v$ the speed of the flow,
and $r$ the typical size of the fireball.  Since $\epsilon/T^4=4.4\pm0.1$
for the QCD plasma at $2T_c$, $s/T^3=5.7\pm0.2$ and $v=c_s\approx1/\sqrt3$
\cite{swagato}, using the AdS/CFT estimate \cite{son} $\eta/s\ge1/4\pi$,
one finds $R\le5.6(rT)$. In fact, on the lattice \cite{sgupta}, one
finds $\eta/s$ which is twice this value, giving
\beq
   R=2.8(rT).
\label{estre}\eeq
This estimate is made at $2T_c\approx 350$ MeV.  However, direct
data on $r$ comes from HBT analysis. At chemical freezeout temperature,
$T\approx175$ MeV, this implies $r=7$ fm.  By using the freezeout
values in the estimate of $R$, one overestimates it, because $\eta$
is espected to be larger at freezeout (otherwise matter would not
freeze out) and $s$ is expected to drop, pushing up the value of
$\eta/s$ \cite{gyulassy}. On the other hand, if we take the temperature
to be $2T_c$, then $r$ would be smaller. So an upper bound on $R$
is obtained by using freezeout values of $r$ and $T$ in eq.\
(\ref{estre}).  This is $R<18$.

If the Bjorken flow approximation is not destroyed by viscosity,
then one could use the estimate $T\propto\tau^{-1/c_s^2} \simeq\tau^{-3}$
and $r\simeq\tau$ to write $rt\propto\tau^{-2}$. If one neglects
any special kinetics that develops at the QCD cross over, and uses
Bjorken flow to push the estimate above back to a time when the
temperature was $2T_c$, then one obtains $R\simeq14$. The main
correction to this comes from the observation on the lattice of the
slowing of sound as one approaches $T_c$ \cite{swagato}. We will
later investigate the stability of Bjorken flow against dissipation.

In non-relativistic dynamics of a liquid, one also has
incompressibility. Low $R$ and incompressibility together cause the
motion to be reversible.  This doesn't happen in heavy-ion collisions.
Write $\eta=\epsilon c_s\lambda$ where $c_s$ is the speed of sound
and $\lambda$ is a typical length characteristic of momentum
transport (which is $c_s\tau/3$ in a non-relativistic gas). Then
\beq
   R=\left(\frac v{c_s}\right)\,\left(\frac r\lambda\right)
    =M\,\left(\frac r\lambda\right)
\label{rereynolds}\eeq
where $M$ is the Mach number. For heavy-ion collisions, $1\le
M\le\sqrt3$.  The lower bound comes from the fact that matter is
expanding into a vacuum so the transverse flow essentially moves
at the speed of sound. The upper bound is for fast particles or
jets travelling through the medium.  Therefore we have flow at small
Reynolds number and large Mach number.

Typically a large Mach number means that compression cannot be
neglected and shocks are possible. A low Reynolds number means
that energy can be dissipated during a shock. So if one has viscous
shock waves in heavy-ion collisions, then the energy of the expansion
of the fireball can be pumped back into reheating, thereby prolonging
the life of the plasma droplet. This scenario has been studied in
\cite{stocker,shuryak,chaudhury}.

But instead of stopping here to follow up on this, we proceed with
dimensional analysis.  We write the Knudsen number
\beq
   K = \frac\lambda r = \frac RM\,.
\label{flouquet}\eeq
Using the values of the Reynolds and Mach numbers presented before,
we find that--- $\lambda/r\approx$0.032--0.055. Another estimate,
taking $r\approx7$ fm and $\lambda\approx0.3$ fm \cite{sgupta}, is
$K\approx0.04$, which is consistent with this.  Hydrodynamics is
the limit of very small Knudsen number. Is the number we obtained
here small in this sense? To understand this, we note that a finite
$K$ means that the isotherms of ideal fluid flow cannot be regarded
as mathematically perfect surfaces of zero thickness, but are smeared
out over 3--6\% of the radius of the fireball. In particular, this
means that the outer skin, from which particles can leave the
fireball, is 10--20\% of the volume if the fireball shape is a
sphere.  If the fireball is elongated, then the evaporative skin
can be a larger fraction of the volume. Thus, in the time a low temperature
isotherm moves half the way towards the center, almost a quarter of the
material from the surface has evaporated.  So the values of $K$
here are not really small enough for ideal hydro, since the hot
system is already pretty close to freezeout.

A very enlightening way in which to consider the strength of dissipative
effects is to compute the damping of sound waves. In the presence of
viscosity, sound instensity is exponentially damped: $\exp(-r/\Gamma_s)$.
The sound attenutation length is given by
\beq
   \Gamma_sT = \frac{4\eta}{3s} \ge \frac1{3\pi},
\label{sal}\eeq
where the last estimate is derived from the bound conjectured in
\cite{son}. At $2T_c$, lattice results would predict $\Gamma_s$ to
be twice this limiting value \cite{sgupta}. With this estimate, one
e-fold decrease in intensity would occur over a distance of less
than 0.4 fm at $2T_c$.  This has strong implications on the
hydrodynamic history of the fireball, since both the transverse
flow and elliptic flow are acoustic in nature, driven
by pressure gradients. Such a strong dissipation of sound implies
that the hydrodynamic prediction of the elliptic flow
must be smaller.

So, next we ask how the dissipation affects the longitudinal flow---
this may be different. Using first order dissipative hydrodynamics
\cite{ll}, and doing a simple order of magnitude estimate of the
relative importance of the entropy current, $s'$, and the entropy
$s$, one finds
\beq
   \frac{s'}{s}
        = \left(\frac\eta s\right)\,\left(\frac{u^2}{Tr^2}\right)
\label{rate}\eeq
Now $u=1$, $T=150$ MeV and $r=7$ fm at freezeout. If we take the
AdS/CFT lower bound, $\eta/s=1/4\pi$,  then $s'/s\approx0.0014$
near freezeout. This small number is an underestimate because
$\eta/S$ has to be larger at freezeout, and the bound is more
accurate at early times, when one has larger $T$ and smaller $r$.
If the entropy production were small at all times, one could just
estimate it for Bjorken flow. Then, as before, $T\propto\tau^{-3}$
and $r\propto\tau$, one would have $s'/s\propto\tau$. This allows
us to push the estimate above back to a time appropriate to the
input value of $\eta/S$, and shows that the entropy production at
earlier times would have been negligible.  This analysis indicates
that the Bjorken solution is stable against first order dissipative
terms even if $\eta/s$ is significantly larger than the limit
conjectured in \cite{son}. This, in turn, lends stability to our
earlier estimate of the Reynolds number.

While we are discussing dissipation, I make a detour to ask how big
is the bulk viscosity, $\zeta$. This is more or less unknown.  In
a perturbative computation it would be very small. General conditions
are discussed for the vanishing of $\zeta$ in \cite{wbook}, where
a result from \cite{tisza} is quoted which says that if the trace
of the stress tensor, $\epsilon-3P=f(\epsilon,n)$ ($P$ is the
pressure and $n$, a conserved particle number), then $\zeta$ vanishes.
This condition fails to hold in many cases, most notably when the
temperature is comparable to masses of the particles or when matter
contains composite particles and energy can be transferred from
translational modes to internal excitations \cite{starinets}. In
any case, a stress tensor with non-vanishing trace is a breakdown
of conformal symmetry, and may exist in any field theory with a
non-vanishing $\beta$-function.  In particular, in QCD near $T_c$
both these conditions may hold.

The bulk viscosity is known for a two component model of matter
where one component has very small mean free time compared to the
other \cite{weinberg}, and gives
\beq
   \frac\zeta\eta = \frac53\left(1-3\cs^2\right)^2,
\label{bulk}\eeq
where $\cs$ is the speed of sound in the medium. Several attempts
to understand the behaviour of the equation of state of QCD near
$T_c$ have effectively involved two component models which are
similiar. Moreover, at least one of the components in such a model
is usually understood to be composite \cite{note1}, and thereby
satisfy one of the conditions for a large $\zeta$. Since the speed
of sound is now known directly from lattice computations \cite{swagato},
one can form an estimate of this ratio. Of course, this would be
correct only if QCD matter were approximately of this kind. On the
other hand, if it is not, having some estimate available may allow
us to design observables that would test or rule out this scenario.
Using the data of \cite{swagato} in conjunction with eq.\ (\ref{bulk}),
we find that $\zeta/\eta=0.45$ at $T=0.9T_c$. At $1.1T_c$, immediately
above the transition, $\zeta/\eta=0.35$. Even at $1.5T_c$ the ratio
is as large as 0.1, but it decreases rapidly above that, being
totally negligible at $2T_c$ and above. We end our detour on the
bulk viscosity with this estimate and return to a summary of the
evolution of the fireball.

The viscous corrections to Bjorken flow can be appproximately absorbed
into a modified stress tensor \cite{teaney}. An interesting dimensionless
number to consider in this case is the ratio between the transverse and
longitudinal ``pressures'',
\beq
   A = \frac{P+2\eta/3\tau}{P-4\eta/3\tau} 
     = \frac{(P/sT) + (2/3)(\eta/s)/\tau T}{(P/sT) - (4/3)(\eta/s)/\tau T}\,.
\label{asym}\eeq
In dissipationless Bjorken hydrodynamics, $A=1$.  In this form one
can use the estimate $P/T^4=1.26\pm0.04$ along with the previous
estimates at $2T_c$. Then one finds $A=1.37$ if one uses the Bjorken
solution to propagate the freezeout value of $\tau T$ backwards to
a time when $T=2T_c$. If instead one uses the freezeout value of
this combination one finds $A=1.28$. The small departure from $A=1$
is another indication of the stability of the Bjorken solution.
This is, of course, one of the bases of the analysis in \cite{teaney}.

The salient qualitative results of our analysis are---
\begin{enumerate}
\item The use of ideal hydrodynamics is incorrect but may not be
  terribly wrong in some ways. The isotherms are smeared out to
  3--6\% of the system size. However, it is bad in other ways;
  for example, the use of the Bjorken formula to extract
  the initial energy density is probably too optimistic. It seems
  possible that the initial energy density is significantly smaller---
  perhaps by 15--25\%. A loss of another 15--25\% by surface emission
  means that perhaps only about half of the Bjorken estimate of the
  energy density is available to the fireball.
\item The transverse flow stalls. The reason is that this is a
  sound wave, driven by the pressure gradient between the core and
  the surface of the fireball. However, the sound attenuation length
  is short, since $R$ is large. Therefore the pressure gradient is unable
  to drive a rarefaction wave efficiently. This increases the lifetime of
  the plasma over that expected from ideal hydrodynamics.
\item This reasoning also applies to differential pressure gradients
  which cause elliptic flow, $v_2$. Thus, the true elliptic flow must
  be at least 20\% smaller than the ideal hydro result \cite{molnar}. In
  that case, the increase of $v_2$ with $s$ \cite{na49} must be re-examined
  to check whether it can be used to extract $\eta/s$.
\item Bjorken flow is fairly stable under the amount of dissipation that
  seems to exist in the QCD plasma. The amount of entropy generated by
  purely logitudinal flow is small. However, pressure isotropy is violated
  by 20--30\% during Bjorken flow. Note however, that this will not
  drive acoustic phenomena.
\item Any pre-equilibrium signal that one gets at the detector comes from
  the outer region of the fireball, and therefore from matter at fairly
  low densities. These signals should be computable in a hadronic model.
\end{enumerate}

We believe that all this is good news.  If ideal hydrodynamics were
a quantitatively good theory of the fireball's evolution, then the
freezeout isotherm would have to be opaque in hadronic channels.
One could not then look past it into the fireball.  All observable
hadronic quantities--- single particle spectra, HBT correlations,
fluctuations of conserved quantities, chemical abundances, would
give direct information only on freezeout conditions, which lie in
the hadronic phase. In this case, the quark-gluon plasma would be
invisible; its existence would have to be deduced by fitting the
equation of state to hydrodynamics--- a notoriously ill-conditioned
problem.  So it is good news that the quark-gluon plasma is not
totally invisible.  One can look into the history of the plasma.
However, this means that the tool to extract physics becomes a
little complicated--- one needs transport equations rather than
hydrodynamics.

I would like to thank Jean-Yves Ollitrault for discussion and a
reading of the paper during a visit to CEA Saclay under the IFCPAR
project 3104-3 called ``Hot and Dense QCD Matter''.



\begin{thebibliography}{99}
\bibitem{note1}
   Very close to $T_c$, lattice computations imply that the assumption that there
   are gluon-like quasi-particles in the plasma is false \cite{saumen2}. The
   conjecture that the plasma is full of coloured composites \cite{shuryak2} would
   give rise to interesting transport phenomena, and can be tested by these means.
\end{thebibliography}
\end{document}